\documentclass[aps,prd,10pt,nofootinbib,OliveGeqsecnum,showpacs,showkeys,superscriptaddress,preprintnumbers,altaffilletter,floatfix,twocolumn]{revtex4-2}

\usepackage[utf8]{inputenc}    
\usepackage[T1]{fontenc}       
\usepackage[english]{babel}    
\usepackage{lmodern}           

\usepackage{amsmath}
\usepackage{multirow}
\usepackage{hyperref}
\usepackage{soul}
\usepackage{graphicx}
\usepackage{dcolumn}
\usepackage{amssymb}
\usepackage{amsfonts}
\usepackage{amsbsy}
\usepackage{color}
\usepackage{rotating}
\usepackage{overpic}
\usepackage{float}
\usepackage{bbold}
\usepackage{enumerate}
\usepackage{dsfont}
\usepackage{mathrsfs}
\usepackage{mathtools}
\usepackage{tensor}
\usepackage{caption}
\usepackage{subcaption}
\usepackage{ragged2e}

\definecolor{amaranth}{rgb}{0.9, 0.17, 0.31}
\definecolor{palatinateblue}{rgb}{0.15, 0.23, 0.89}
\definecolor{brightpink}{rgb}{1.0, 0.0, 0.5}

\hypersetup{
    linktoc=all,
    colorlinks=true,
    linkcolor={palatinateblue},
    citecolor={brightpink},
    urlcolor={amaranth}
}

\graphicspath{ {images/} }

\begin{document}

\title{Speeding up the Universe with a generalised axion-like potential}

\author{Carlos G. Boiza}
\email{carlos.garciab@ehu.eus}
\affiliation{Department of Physics \& EHU Quantum Center, University of the Basque Country UPV/EHU, P.O. Box 644, 48080 Bilbao, Spain}
\author{Mariam Bouhmadi-López}
\email{mariam.bouhmadi@ehu.eus}
\affiliation{IKERBASQUE, Basque Foundation for Science, 48011, Bilbao, Spain}
\affiliation{Department of Physics  \& EHU Quantum Center, University of the Basque Country UPV/EHU, P.O. Box 644, 48080 Bilbao, Spain}

\begin{abstract} 
Understanding the late-time acceleration of the Universe is one of the major challenges in cosmology today. In this paper, we present a new scalar field model corresponding to a generalised axion-like potential. In fact, this model can be framed as a quintessence model based on physically motivated considerations. This potential is capable of alleviating the coincidence problem through a tracking regime. We will as well prove that this potential allows for a late-time acceleration period induced by an effective cosmological constant, which is reached without fine-tuning the initial conditions of the scalar field. In our model, the generalised axion field fuels the late-time acceleration of the Universe rather than fuelling an early dark energy era. Additionally, we will show how the late-time transition to dark energy dominance could be favoured in this model, since the density parameter of the scalar field will rapidly grow in the late phase of the tracking regime.

\end{abstract}

\maketitle

\renewcommand{\tocname}{Index}


\section{Introduction}\label{intro}

The origin of the accelerated expansion we observe in the Universe \cite{SupernovaSearchTeam:1998fmf,SupernovaCosmologyProject:1998vns,Bahcall:1999xn} remains unknown from a fundamental point of view. The simplest model capable of describing this accelerated expansion is constructed by adding a positive cosmological constant to Einstein's equations, known as the $\Lambda$CDM model. In this model, the energy density of dark energy remains constant over time. Although this model can explain most cosmological observations, it presents several problems both at fundamental and experimental levels. How is it possible that dark energy and dark matter, which seem to dominate the current stage of the Universe, are comparable precisely today, and how is this fact affected by specific initial conditions? This is known as the coincidence problem \cite{Carroll:2000fy,Padmanabhan:2002ji}. Additionally, the cosmological constant suffers from a fine-tuning problem, as it introduces a new energy scale, $\rho_{\Lambda}\approx10^{-47}\,\textrm{GeV}^{\,4}$, that is very small compared to other scales in particle physics \cite{Weinberg:1988cp,Sahni:1999gb}. On the other hand, there is a tension when it comes to extracting the current rate of expansion of the Universe: direct measurements favour a higher value of $H_0$ compared to the value obtained through indirect measurements. There is also a tension in the estimation of matter clustering as measured through $\sigma_8$: measurements of that quantity from late-time cosmological observations differ from those deduced from measurements of the early Universe. The Hubble tension and the $\sigma_8$ tension are open problems in cosmology that remain unsolved \cite{DiValentino:2021izs,Poulin:2018cxd,Kamionkowski:2022pkx,DiValentino:2020vvd,Perivolaropoulos:2021jda,Abdalla:2022yfr}. 

For all these reasons, new models beyond $\Lambda$CDM have been proposed to explain the accelerated expansion of the late Universe. There are essentially two main paths to tackle this issue.  On the one hand, one can invoke a dark energy component. The simplest case beyond the $\Lambda$CDM paradigm is to invoke a perfect fluid, which can be (i) within the \textit{canonical} regime, i.e., an equation of state, $w$, such that $w>-1$ (for example, the Chaplygin gas \cite{Kamenshchik:2001cp}), or (ii)  in the phantom regime, i.e., $w<-1$ (see, for example, \cite{Albarran:2016mdu}). On the other hand, one can consider an extended theory of gravity, which can be constructed, for example, within  metric modified gravity, where the metric is the fundamental variable; Palatini modified gravity, where the metric and connection are treated as independent variables; and non-metric modified gravity, in which the metric does not satisfy the compatibility condition and the connection is not derived from it (for an extended review on these topics, please see \cite{CANTATA:2021asi}). In this work, we focus on a dark energy model built from a canonical scalar field minimally coupled to gravity, known as quintessence \cite{Ratra:1987rm}. Non-canonical scalar fields, which can give rise to phantom behaviours, and quintom models describing the crossing of the phantom divide, have also been studied \cite{Caldwell:1999ew,Cai:2009zp}. More general scalar field theories can be described within the framework of the Horndeski model (cf. \cite{Kobayashi:2019hrl} for a review).

Quintessence models allow for a dynamical explanation of the dark energy component with an equation of state equal to or greater than $-1$. Although DESI BAO 2024 data \cite{DESI:2024mwx} alone do not favour a dynamical dark energy sector, their combination with other cosmological data does. In fact, an equation of state today greater than $-1$ seems to be favoured, but it is too early to draw conclusions, and all possibilities are still open. Quintessence models have proven useful in alleviating the coincidence problem through scaling solutions \cite{Wetterich:1987fm,Ferreira:1997hj}. In \cite{Caldwell:2005tm}, the authors divided quintessence models into two categories: thawing and freezing (see also \cite{Clemson:2008ua,Chiba:2012cb,Pantazis:2016nky}). Thawing potentials are initially frozen at $w\approx-1$, but since $w'>0$, i.e., the equation of state is a growing function of time, the cosmological constant behaviour is lost at some point. Examples of potentials with these characteristics are power-law potentials $\phi^n$ with positive power, i.e., $n>0$, thoroughly studied in the literature (see, for example, \cite{Alho:2015cza}), and  axion-like potentials $[1-\cos{(\phi/\eta)}]^n$, with $n>0$. The latter has been used to alleviate the coincidence problem in the context of a string axiverse \cite{Kamionkowski:2014zda} (see also \cite{Emami:2016mrt}) and has also been proposed as a candidate for early dark energy capable of alleviating the Hubble tension \cite{Poulin:2018cxd,Kamionkowski:2022pkx}. The use of this potential is also found in Natural Inflation \cite{Freese:1990rb}. The potential has also been invoked in the context of ultralight dark matter, when oscillations around the minimum are happening (for a detailed analysis of wave dark matter, see \cite{Hui:2016ltb}). Ultralight axions have been proved to be a feasible dark matter candidate, not only from a pure theoretical viewpoint, but also by using cosmological simulations \cite{Schive:2014dra,Schive:2014hza,Mocz:2019uyd}. For further details on axions applied to cosmology, see \cite{Ng:2000di,Hammer:2020dqp,Marsh:2015xka,Chakraborty:2021vcr}. On the other hand, freezing quintessence models initially have $w>-1$ and $w'<0$ and evolve asymptotically to $w\rightarrow-1$, $w'\rightarrow0$. This kind of potentials have been used to alleviate the coincidence problem through tracking behaviours \cite{Zlatev:1998tr,Steinhardt:1999nw}. The simplest case of a freezing potential with tracking is given by a power-law potential $\phi^n$ with a negative power, i.e., $n<0$ \cite{Zlatev:1998tr,Liddle:1998xm,Steinhardt:1999nw,delaMacorra:2001ay}.

In this work, we construct a quintessence model inspired by the structure of axion-like potentials. In particular, we will show that a generalisation of the usual axion-like form to negative powers, i.e., $n<0$, arises naturally within the zoo of freezing scalar field potentials. Although this model is not derived from a fundamental high-energy theory such as string theory or spontaneous symmetry breaking, its form is motivated by phenomenological considerations and dynamical requirements relevant to late-time cosmology. This potential is capable of alleviating the coincidence problem through a tracking regime. It will be shown that this potential will allow for a late-time acceleration period caused by an effective cosmological constant, which is reached without fine-tuning the initial conditions of the scalar field. In our model, the generalised axion field fuels the late-time acceleration of the Universe rather than fuelling an early dark energy era. In addition, we will see how the late-time transition to dark energy dominance could be favoured in this model through a specific mechanism that makes the density parameter of the scalar field rapidly grow in the late phase of the tracking regime.

The paper is organised as follows. In Section \ref{sfdem}, we will review the general aspects of quintessence models, focusing on the dynamical aspects and on how a tracking behaviour can alleviate the coincidence problem. Then, in Section \ref{pmalpltc}, we will propose a new quintessence model based on physically motivated considerations. We will find that the model is a generalisation of the axion-like potential, which, in our case, will fuel the late-time acceleration. This is due to the appearance of a late-time attractor in the system, which behaves like an effective cosmological constant and is reached close to the present time without fine-tuning the initial conditions of the field. In Section \ref{dsam}, we will see how the dynamical properties introduced in Section \ref{sfdem} work in our specific models and how tracking trajectories appear, alleviating the coincidence problem. We will complement the theoretical considerations with several numerical calculations to see how the dynamics of the scalar field behaves in our particular model. Finally, in Section \ref{conclusions}, we will conclude with some final remarks on the work. In a companion paper \cite{Chiang:2025qxg}, we test this model against a broad set of current cosmological observations, including measurements of the cosmic microwave background, baryon acoustic oscillations, supernovae, the Hubble constant at low redshift, and large-scale structure.

\section{Scalar field dark energy models}\label{sfdem}

In this section, we review the cosmological evolution of a canonical scalar field and the dynamical system analysis for a Friedmann–Lemaître–Robertson–Walker (FLRW) universe filled by a scalar field and a perfect fluid. In particular, we review under which conditions tracking behaviour of a scalar field, within this setup, can show up. This will be extremely useful for the analysis we carry for our model in Section \ref{dsam}.

\subsection{Equations of motion}\label{csfflrw}

Our starting point will be the following action:

\begin{equation}\label{scalaraction}
   S=\int d^4x\sqrt{-g}\left(\frac{R}{2k^2}+L_m+L_{\phi}\right),
\end{equation}
where $k^2=8\pi G=1/M_P^2$ (we use natural units $\hbar=c=1$ throughout this work and $M_P\approx2.435\times10^{18}\,\textrm{GeV}$ is the reduced Planck mass), $R$ is the scalar curvature, $L_m$ is the Lagrangian corresponding to matter and $L_{\phi}$ is the Lagrangian of a minimally coupled scalar field $\phi$:

\begin{equation}\label{canonicallagrangian}
   L_{\phi}=-\frac{1}{2}g^{\mu\nu}\partial_{\mu}\phi\partial_{\nu}\phi-V(\phi).
\end{equation}
In the former expression, $V(\phi)$ is the potential of the scalar field, which we assume to be positive.

Minimising the action \eqref{scalaraction} with respect to variations of the scalar field $\phi$, we obtain the classical Klein-Gordon equation given by

\begin{equation}\label{kgeq}
   \Box\phi-V_{,\phi}=0,
\end{equation}
where $V_{,\phi}$ is the derivative of $V$ with respect to $\phi$.

Analogously, minimising the action \eqref{scalaraction} with respect to variations of the metric, we obtain Einstein field equations:

\begin{equation}\label{einsteineq}
   G_{\mu\nu}=k^2(T^{m}_{\mu\nu}+T^{\phi}_{\mu\nu}),
\end{equation}
where $G_{\mu\nu}=R_{\mu\nu}-g_{\mu\nu}R/2$ is the Einstein tensor, $T^{m}_{\mu\nu}$ is the energy-momentum tensor of the matter background and $T^{\phi}_{\mu\nu}$ is the energy-momentum tensor of the scalar field given by

\begin{equation}\label{scalarsetensor}
   T^{\phi}_{\mu\nu}=\partial_{\mu}\phi\partial_{\nu}\phi-\frac{1}{2}g_{\mu\nu}g^{\alpha\beta}\partial_{\alpha}\phi\partial_{\beta}\phi-g_{\mu\nu}V(\phi).
\end{equation}

From now on, we assume a homogeneous and isotropic spatially flat universe, i.e.,  whose metric is described by the FLRW metric:

\begin{equation}\label{rwmetric}
   ds^2=-dt^2+a^2(t)d\bar{x}^2.
\end{equation}

Under these assumptions, \eqref{kgeq} reduces to

\begin{equation}\label{rwkgeq}
   \ddot{\phi}+3H\dot{\phi}+V_{,\phi}=0,
\end{equation}
where a dot stands for derivative with respect to the cosmic time $t$.

Considering the background matter component to be a barotropic fluid with a constant equation of state given by $w=\rho/p$, the only two independent equations resulting from \eqref{einsteineq} are the Friedmann and Raychaudhuri equations:

\begin{equation}\label{friedmanneq}
   H^2=\frac{k^2}{3}\left[\rho+\frac{1}{2}\dot{\phi}^2+V(\phi)\right],
\end{equation}

\begin{equation}\label{raychaudhurieq}
   \dot{H}=-\frac{k^2}{2}\left[(1+w)\rho+\dot{\phi}^2\right].
\end{equation}

We remind that  the energy density and the pressure of the scalar field can be written as 

\begin{equation}\label{denspresscf}
   \rho_{\phi}=\frac{1}{2}\dot{\phi}^2+V(\phi), \quad\quad\quad  p_{\phi}=\frac{1}{2}\dot{\phi}^2-V(\phi),
\end{equation}
which are, of course, consistent with \eqref{friedmanneq} and \eqref{raychaudhurieq} and easily deduced from \eqref{scalarsetensor}.

\subsection{Dynamical system}\label{ds}

In order to study the evolution of the system given by \eqref{rwkgeq}, \eqref{friedmanneq} and \eqref{raychaudhurieq}, it is convenient to introduce a new set of variables defined as \cite{Copeland:1997et}

\begin{equation}\label{xexpression}
   x=\frac{k\dot{\phi}}{\sqrt{6}H},
\end{equation}

\begin{equation}\label{yexpression}
   y=\frac{k\sqrt{V}}{\sqrt{3}H}.
\end{equation}
Differentiating with respect to $\log{(a/a_0)}$, where $a_0$ is the current value of the scale factor, and using \eqref{rwkgeq}, \eqref{friedmanneq} and \eqref{raychaudhurieq}, we obtain the following dynamical system:

\begin{eqnarray}\label{xeq}
   x'=-\frac{3}{2}\left[2x+(w-1)x^3+x(w+1)(y^2-1)-\sqrt{\frac{2}{3}}\lambda y^2\right],\nonumber\\
\end{eqnarray}

\begin{eqnarray}\label{yeq}
   y'=-\frac{3}{2}y\left[(w-1)x^2+(w+1)(y^2-1)+\sqrt{\frac{2}{3}}\lambda x\right],\nonumber\\
\end{eqnarray}
where the variable $\lambda$ has been defined as 

\begin{equation}\label{lambdadef}
   \lambda=-\frac{V_{,\phi}}{kV}.
\end{equation}
Since $\lambda$ depends on $\phi$, the dynamical system is still not closed. In order to get a closed one, it is possible to treat $\lambda$ as another dynamical variable. With this in mind, after differentiating $\lambda$ with respect to $\log{(a/a_0)}$ we obtain \cite{Steinhardt:1999nw,delaMacorra:1999ff,Ng:2001hs}

\begin{equation}\label{lambdagammaeq}
   \lambda'=-\sqrt{6}(\Gamma-1)\lambda^2x,
\end{equation}
where $\Gamma$ is defined as 

\begin{equation}\label{gammadef}
   \Gamma=\frac{VV_{,\phi\phi}}{V_{,\phi}^2},
\end{equation}
and $V_{,\phi\phi}$ is the second derivative of $V$ with respect to $\phi$. If $\lambda(\phi)$ is invertible, it is possible to express $\phi$ as a function of $\lambda$: $\phi(\lambda)$. In this case, $\Gamma$ can be expressed in terms of $\lambda$ and \eqref{xeq}, \eqref{yeq} and \eqref{lambdagammaeq} form, together, a closed dynamical system. For simplicity, we write the evolution equation of $\lambda$ as

\begin{equation}\label{lambdafeq}
   \lambda'=-\sqrt{6}f(\lambda)x,
\end{equation}
where we have defined $f(\lambda)=\lambda^2[\Gamma(\lambda)-1]$.

In terms of the new variables, the Friedmann equation \eqref{friedmanneq} reduces to $x^2+y^2+\Omega_m=1$, where $\Omega_m=k^2\rho/(3H^2)$ is the relative energy density of matter, which is always positive. This implies the constraint $0\leq x^2+y^2\leq 1$. 

It is useful to write the cosmological parameters in terms of the dynamical variables. For example, the relative energy density of the scalar field is given by

\begin{equation}\label{omegaphi}
   \Omega_{\phi}=\frac{k^2\rho_{\phi}}{3H^2}=x^2+y^2,
\end{equation}
and the scalar field and the total equations of state can be written, respectively, as

\begin{equation}\label{fieldeos}
   w_{\phi}=\frac{p_{\phi}}{\rho_{\phi}}=\frac{x^2-y^2}{x^2+y^2},
\end{equation}

\begin{equation}\label{effeos}
   w_{\textrm{eff}}=\frac{p+p_{\phi}}{\rho+\rho_{\phi}}=x^2-y^2+w(1-x^2-y^2).
\end{equation}

In Tab. \ref{tab:fixedpointsgeneral}, we have classified all the fixed points which could appear in the finite regime of the phase space assuming a general potential $V$ and a finite $f$. In order to study the system at the infinite points $\lambda\rightarrow\pm\infty$, assuming that $\Gamma$ remains finite at these points, a compactification of the variable $\lambda$ as the one proposed in \cite{Ng:2001hs} must be performed (cf. also \cite{Bouhmadi-Lopez:2016dzw,BorislavovVasilev:2022gpp}).

\begin{table}[t]
    \centering
    \begin{tabular}{||c|c|c|c||}
    \hline
       \:\:Point\:\: & $x$ & $y$ & $\lambda$ \\
       \hline
       \hline
       A  & $0$ & $0$ & \:\:Any\:\:\\
       \:\:\:B$_{\pm}$  & $\pm1$ & $0$ & $\lambda_*$\\
       C  & $\:\:\sqrt{3}(1+w)/(\sqrt{2}\lambda_*)\:\:$ & $\:\:\sqrt{3(1-w^2)/(2\lambda_*^2)}\:\:$ & $\lambda_*$\\
       D  & $\lambda_*/\sqrt{6}$ & $\sqrt{1-\lambda_*^2/6}$ & $\lambda_*$\\
       E  & $0$ & $1$ & $0$\\
       \hline
    \end{tabular}
    \caption{\justifying{Fixed points of the system given by \eqref{xeq}, \eqref{yeq} and \eqref{lambdafeq}. $\lambda_*$ is any zero of the function $f(\lambda)$. This table is taken from \cite{Bahamonde:2017ize} for the sake of clarification of our discussion. This table is valid for finite phase space variables such that $f$ remains as well finite.}}
    \label{tab:fixedpointsgeneral}
\end{table}

For our later analysis, it is convenient to highlight some of the fixed points deeply analysed in \cite{Bahamonde:2017ize}. Among all of them, only two can lead to an accelerated expanding universe: points D and E. The point D only exists if $\lambda_*^2<6$ and, since $w_{\textrm{eff}}=\lambda_*^2/3-1$ at this point, it corresponds to an accelerated expanding universe if $\lambda_*^2<2$. The point E always exists independently of the potential chosen. It corresponds to a de Sitter-like universe with $w_{\textrm{eff}}=-1$. Its stability depends on the sign of the the function $f$ evaluated at $\lambda=0$: it is stable if $f(0)>0$ and unstable if $f(0)<0$. If $f(0)=0$, the centre manifold theorem must be applied in order to determine its stability. The point C is also important because it leads to a scaling behaviour where the proportion between matter and dark energy could be comparable: $\Omega_{\phi}=3(1+w)/\lambda_*^2$ and $\Omega_m=1-3(1+w)/\lambda_*^2$. Therefore, its appearance in the dynamical system as an attractor can alleviate the coincidence problem. The point C exists if $\lambda_*^2\geq 3(1+w)$. However, as $w_{\phi}=w_{\textrm{eff}}=w$ at this fixed point, it can not correspond to an accelerated expanding universe, as $w$ is positive for dark matter and radiation. In addition, a dark energy equation of sate far enough from the cosmological constant one ($w_{v}=-1$) could be problematic when explaining the matter structure formation. Several models transitioning from a scaling point C to a late-time accelerated expanding universe properly described by points D or E have been constructed (see \cite{Barreiro:1999zs,Zhou:2007xp}).

\subsection{Tracking behaviour}\label{tb}

In addition to the appearance of the point C in the dynamical system, which leads to a scaling behaviour, some scalar fields go through an intermediate regime that can alleviate the coincidence problem: the tracking regime. In this phase of the evolution, the equation of state of the scalar field, $w_{\phi}$, remains almost constant and the information about the initial conditions is lost because of the appearance of an attractor in the dynamical system. In \cite{Ng:2001hs}, the authors studied the regime $\lambda\gg1$ and, assuming $\Gamma\approx const.$, they found the existence of what they called ``instantaneous critical points". At these points, the variables $x$ and $y$ satisfy the linear relation $y/x=\sqrt{2}(1+w_{\phi})^{-1/2}-1$, where $w_{\phi}$ is almost constant and is related to $\Gamma$ through the relation

\begin{equation}\label{sfeosgamma}
   w_{\phi}=\frac{w-2(\Gamma-1)}{1+2(\Gamma-1)}.
\end{equation}

The stability conditions of the ``instantaneous critical points" were also analysed in \cite{Ng:2001hs}. If these points are attractors, the system will tend to the tracking regime independently of the initial conditions, alleviating the coincidence problem. In \cite{Steinhardt:1999nw}, the authors proposed a theorem encompassing the conditions that a potential must satisfy in order to have tracking behaviour. Those conditions are the same leading to attractors corresponding to ``instantaneous critical points":

\begin{itemize}
    \item Tracking behaviour with $w_{\phi}<w$ occurs for any potential in which $\Gamma>1$ and $\Gamma\approx const.$ in the regime $\lambda\gg1$.
    
    \item Tracking behaviour with $w<w_{\phi}<(1+w)/2$ occurs for any potential in which $1-(1-w)/(6+2w)<\Gamma<1$ and $\Gamma\approx const.$ in the regime $\lambda\gg1$.
\end{itemize}
The second type of tracking behaviour implies a period where the dark energy equation of state is far from a cosmological constant ($w_{v}=-1$) and the matter structure formation could be difficult to explain (the same problem as the one mentioned in the previous subsection when the point C appears in the dynamical system).

In \cite{Steinhardt:1999nw}, the authors also suggested $\Gamma$ functions that increase with time in order to naturally explain the dominance of dark energy in the Universe today rather than at earlier epochs. Since $\Omega_{\phi}\propto t^{P}$ is approximately satisfied, where $P$ is a function of $\Gamma$ given by \footnote{We are assuming a matter dominated tracking regime in which $w_{\phi}$ remains almost constant and is given by \eqref{sfeosgamma}. In that case, the density parameter of the scalar field scales as $\Omega_{\phi}\propto a^{-3(w_{\phi}-w)}$, where $w$ is the matter equation of state. Substituting \eqref{sfeosgamma} into the former expression and remembering that $a\propto t^{2/[3(1+w)]}$ in a matter dominated universe, we finally obtain the time dependence of the density parameter encompassed in \eqref{growgamma}.}

\begin{equation}\label{growgamma}
   P=\frac{4(\Gamma-1)}{1+2(\Gamma-1)},
\end{equation}
we see how the growth of $\Omega_{\phi}$ depends on $\Gamma$. If $\Gamma$ is constant, as in the case of an exponential potential or an inverse power-law potential \footnote{Direct power-law potentials have also a constant $\Gamma$ function (cf. \eqref{gammadef}). In fact, for a general power-law potential $\phi^n$, the $\Gamma$ function is given by $\Gamma=(n-1)/n$. We exclude direct power-law potentials ($n>0$) because they do not satisfy $\Gamma>1$, one of the necessary conditions in order to have physically viable tracking behaviour. On the contrary, inverse power-law potentials ($n<0$) satisfy all the necessary conditions.}, $\Omega_{\phi}$ will grow as a power law function (for an exponential potential, it will remain constant since $\Gamma=1$ and $P=0$). If $P$ increases at late-time in the final stage of the tracking regime, as could happen if we choose $\Gamma>1$ which increases at late-time, $\Omega_{\phi}$ will grow more rapidly. This will cause a quick transition from matter to dark energy dominance, naturally explaining the dominance of the latter today rather than earlier.

\section{A physically motivated generalised axion-like potential for late-time cosmology}\label{pmalpltc}

In this section, we construct a new quintessence model with interesting properties similar to those discussed in the previous section.

Several scalar potentials leading to a broad range of $\Gamma$ functions have been proposed and well studied in the literature \cite{Barreiro:1999zs,Jarv:2004uk,Li:2005ay,Ng:2000di,Urena-Lopez:2011gxx,Gong:2014dia,Ng:2001hs,Fang:2008fw,Roy:2013wqa,Garcia-Salcedo:2015ora,Paliathanasis:2015gga,Matos:2009hf,Zhou:2007xp,Clemson:2008ua}. In this work, we first propose a new $\Gamma$ function and then follow the approach proposed in \cite{Zhou:2007xp} to construct the scalar potential associated to it. The $\Gamma$ function suggested in our work is given by

\begin{equation}\label{gammaprop}
   \Gamma=1+\alpha+\frac{\beta}{\lambda^2},
\end{equation}
where $\alpha$ and $\beta$ are positive dimensionless constants. As we will explain in the next section, this choice of $\Gamma$ satisfies some of the crucial properties to alleviate the coincidence problem and to naturally explain the dominance of dark energy today. On the one hand, $\Gamma$ satisfies the tracking theorem proposed in \cite{Steinhardt:1999nw} and discussed in the previous section since $\Gamma>1$ if $\alpha>0$, $\beta>0$ and $\Gamma\approx const.$ in the regime $\lambda\gg1$. Therefore, we ensure a tracking behaviour under those conditions. On the other hand, if the system evolves such that $\lambda$ approaches today smaller values, this would imply larger values of $\Gamma$, resulting in a natural dark energy dominance at present (and not at higher redshift).

In \cite{Zhou:2007xp}, the author defines $h=V_{,\phi}$ and solves the differential equation \footnote{Note that this is only a fancy way of rewriting the definition of $\Gamma$ given by the expression \eqref{gammadef}. Applying the chain rule, $dh/dV=h_{,\phi}/V_{,\phi}$\:, and noting that $h_{,\phi}=V_{,\phi\phi}$\:, we automatically recover the expression \eqref{gammadef}.}

\begin{equation}\label{difderpot}
   \frac{dh}{dV}=\frac{1}{h}F(V,h),
\end{equation}
where $F=(h^2/V)\Gamma$. In our model, the function $F(V,h)$ is given by

\begin{equation}\label{fgrandmodel}
   F(V,h)=(1+\alpha)\frac{h^2}{V}+\beta k^2V.
\end{equation}
We now solve \eqref{difderpot}. We find the next expression for $h$:

\begin{equation}\label{hmodel}
   h(V)=\pm V\left(c_1V^{2\alpha}-\frac{\beta k^2}{\alpha}\right)^{1/2},
\end{equation}
where $c_1$ is an integration constant with dimensions of $L^{2(1+4\alpha)}$. We can now solve the differential equation $V_{,\phi}=h(V(\phi))$ to obtain the expression of the potential:

\begin{eqnarray}\label{scalarpotentialmodel1}
   &&V(\phi)=\left(\pm\sqrt{\frac{2\beta}{\alpha c_1}}k\right)^{1/\alpha}\nonumber\\
   &&\,\,\,\,\,\,\,\,\,\,\times\left[1+\cos{\left(2\alpha c_2\sqrt{\beta}k+2\sqrt{\alpha\beta}k\phi\right)}\right]^{-1/2\alpha},
\end{eqnarray}
where $c_2$ is another integration constant with dimensions of $L^{-1}$. For the potential to be real valued, we choose real integration constants and restrict the analysis to the $+$ sign solution. Writing $\alpha$ and $\beta$ in terms of two new parameters $n$ and $\eta$ as

\begin{equation}\label{netavariables}
   \alpha=\frac{1}{2n}, \quad \beta=\frac{n}{2k^2\eta^2},
\end{equation}
where $n>0$ in order to preserve the sign of $\alpha$ and $\beta$, the potential \eqref{scalarpotentialmodel1} reads as

\begin{equation}\label{scalarpotentialmodel2}
   V(\phi)=\left(\frac{2n^{2}}{c_1\eta^{2}}\right)^{n}\left[1+\cos{\left(\frac{c_2}{\sqrt{2n}\eta}+\frac{\phi}{\eta}\right)}\right]^{-n}.
\end{equation}
We note that, in order for the constants $\alpha$ and $\beta$ to remain dimensionless, $n$ must also be dimensionless and $\eta$ must have dimensions of $L^{-1}$. We fix the integration constant $c_2$ in order to have a minimum at $\phi/\eta=\pi$: we choose $c_2=\sqrt{2n}\pi\eta$, consistent with the dimension analysis. We also restrict the domain of the potential to $(0,2\pi\eta)$. We additionally redefine the integration constant $c_1$ in order to simplify the expression \eqref{scalarpotentialmodel2} by introducing a new parameter $\Lambda$ with dimensions of $L^{-1}$: $c_1=2n^2/(\eta^2\Lambda^{4/n})$. Taking everything into account, the expression \eqref{scalarpotentialmodel2} written in terms of the parameters $n$, $\eta$ and $\Lambda$ reduces to

\begin{equation}\label{axionscalarpotential}
   V(\phi)=\Lambda^4\left[1-\cos{(\phi/\eta)}\right]^{-n}.
\end{equation}
This is the expression of the potential that we will use throughout this work. Since $n>0$, this expression can be seen as a generalisation of the commonly used axion-like potential $V(\phi)=\Lambda^4[1-\cos{(\phi/\eta)}]^n$ to negative values of $n$. For this reason, from now on we will refer to \eqref{axionscalarpotential} as the \textit{generalised axion-like potential}. Note that the method proposed in \cite{Zhou:2007xp} and followed here allows us to naturally obtain the generalised axion-like potential from a physically motivated $\Gamma$. In this scheme, the parameter $\Lambda$, which is related to the late-time cosmological constant as we will see below, emerges as an integration constant that depends on the initial conditions assumed. Similar conclusions have been found in interacting vacuum models \cite{Wands:2012vg}. Unlike the QCD axion potential, which is tied to the strong CP problem, our generalised axion-like potential is employed solely as a late-time dynamical component within a cosmological context. Although the potential derived in this work does not emerge from a UV-complete theory, it is motivated by the desired dynamical properties of the scalar field and aligns with the effective field theory approach widely used in dark energy phenomenology.

\begin{figure}[t]
\centering
\includegraphics[scale=0.65]{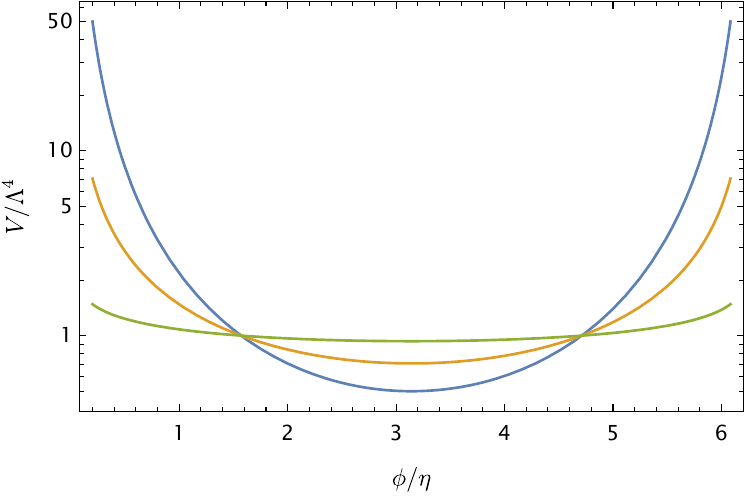} 
\caption{\justifying{This figure shows the shape of the potential $V(\phi)$ given by \eqref{axionscalarpotential}, normalised to $\Lambda^4$. Each colour represents a different value of the parameter $n$: $n=1$ (blue), $n=0.5$ (yellow), $n=0.1$ (green). The potential is flatter for smaller values of $n$, tending to a cosmological constant flat potential in the limit $n\rightarrow0$.}}
\label{fig:potentialshape}
\end{figure}

The potential \eqref{axionscalarpotential} has the interesting property that it allows late-time accelerated expansion through an effective cosmological constant. The minimum value of the commonly used axion-like potential $V(\phi)=\Lambda^4[1-\cos{(\phi/\eta)}]^n$ with $n>0$ is zero, and to construct dark energy models one has to consider regions of the potential near to the maximum where the equation of state could be less than $-1/3$. In \cite{Kamionkowski:2014zda,Emami:2016mrt}, the authors have considered this kind of models and have shown how they could alleviate the coincidence problem through the string axiverse. In this work, we consider the potential \eqref{axionscalarpotential}. The shape of the potential is represented in Fig. \ref{fig:potentialshape} for different values of $n$. We see how the smaller the value of $n$ is, the flatter the potential gets. In fact, in the limit $n\rightarrow 0$ we recover the cosmological constant. If we expand the potential \eqref{axionscalarpotential} around the minimum, we obtain

\begin{equation}\label{minimumexpansion}
   V(\phi)\approx\frac{\Lambda_{\textrm{eff}}}{k^2}+\frac{1}{2}m^2(\phi-\pi\eta)^2,
\end{equation}
where $\Lambda_{\textrm{eff}}=k^2\Lambda^4/2^n$ and $m^2=n\Lambda^4/(2^{n+1}\eta^2)$. In addition to the mass term, we find a non vanishing effective cosmological constant, which is responsible of the late-time accelerated expansion period when the field approaches the minimum of the potential. The values of these terms depend on the parameters of the theory.

We can rewrite \eqref{gammaprop} in terms of the new set of variables $n$, $\eta$ and $\Lambda$:

\begin{equation}\label{gammamodel}
   \Gamma=1+\frac{1}{2n}+\frac{n}{2k^2\eta^2\lambda^2}.
\end{equation}
Calculating $\lambda$ for our model, we find its dependence on $\phi$:

\begin{equation}\label{lambdaphimodel}
   \lambda(\phi)=\frac{n}{k\eta}\frac{\sin{(\phi/\eta)}}{1-\cos{(\phi/\eta)}}.
\end{equation}

Note that, for a critical density today of $\rho_c\sim10^{-47}\,\textrm{GeV}^{\,4}$ ($H_0\sim10^{-33}\,\textrm{eV}$), if the Universe is dominated by the scalar field rolling near the minimum of the potential and neglecting its kinetic term, we have $\rho\sim\Lambda_{\textrm{eff}}/k^2\sim\rho_c\sim10^{-47}\,\textrm{GeV}^{\,4}$, as can be deduced from \eqref{minimumexpansion}. Under these conditions, $\Lambda^4/2^n$ is fixed: $\Lambda^4/2^n\sim10^{-47}\,\textrm{GeV}^{\,4}$. For values of $n$ between $0$ and $1$, we have $\Lambda\sim10^{-3}\,\textrm{eV}$. Moreover, in order for the scalar field to be rolling close to the minimum by today \footnote{In the early Universe, the friction term of \eqref{rwkgeq} is large enough to maintain the scalar field far from the minimum of the potential because $H$ is large. As the Universe evolves, the Hubble expansion rate decreases and the friction term becomes subdominant when $m\sim H$. At this point, the scalar field rolls toward the minimum.}, we need $m\sim H_0\sim10^{-33}\,\textrm{eV}$, which imposes $\eta^2/n\sim10^{37}\,\textrm{GeV}^{\,2}$. With this, for $n\sim 1$, we obtain an axion decay constant approximately equal to the reduced Planck mass: $\eta\sim M_P$. 

In this work, we fix the value $\eta$ to $\eta=1/k=M_P$, and we study two specific cases, corresponding to $n=1$ and $n=0.1$. The axion decay constant is usually restricted to values close to the reduced Planck mass. In \cite{Hui:2016ltb,Marsh:2015xka}, further discussions about the value of this parameter in axion models can be found. Note that, for the chosen value of $\eta$, the model excludes super-Planckian values of the field $\phi$. After fixing $\eta$ and $n$, there is only one free parameter left to adjust: $\Lambda$. We fix it so that we approximately obtain the measured density parameters today: $\Omega_{m0}=0.3$ and $\Omega_{r0}=8\times10^{-5}$ ($\Omega_{\phi0}$ is related to the other density parameters through $\Omega_{\phi0}=1-\Omega_{m0}-\Omega_{r0}$). For $n=0.1$, we find $\Lambda\approx2.102\times10^{-3}\,\textrm{eV}$, which gives us the particle mass $m\approx3.920\times10^{-34}\,\textrm{eV}$. For $n=1$, we find $\Lambda\approx2.177\times10^{-3}\,\textrm{eV}$, which gives us $m\approx9.725\times10^{-34}\,\textrm{eV}$. We have assumed a critical density today of $\rho_c=3H_0^2/k^2\approx3.668\times10^{-47}\,\textrm{GeV}^{\,4}$. All the choices are consistent with the qualitative analysis made in the previous paragraph.

\begin{figure}[t]
\centering
\includegraphics[scale=0.65]{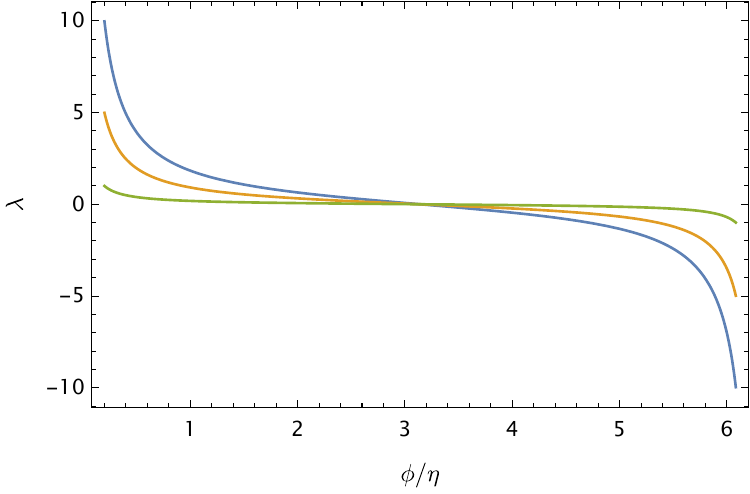} 
\caption{\justifying{This figure shows the behaviour of $\lambda$ as a function of $\phi$ (cf. \eqref{lambdaphimodel}) for different values of the parameter $n$: $n=1$ (blue), $n=0.5$ (yellow), $n=0.1$ (green). The function tends to the constant value $0$ as $n\rightarrow0$, as expected for a cosmological constant.}}
\label{fig:lambdaphimodel}
\end{figure}

The function $\lambda(\phi)$ can take any value in $(-\infty,+\infty)$ (cf. Fig. \ref{fig:lambdaphimodel}). It has an inflection point at $\phi/\eta=\pi$, where the minimum of the potential occurs, as can be seen in Fig. \ref{fig:potentialshape}. The potential diverges in the limits $\phi/\eta\rightarrow0$ and $\phi/\eta\rightarrow2\pi$, where its derivative diverges as well (see Fig. \ref{fig:potentialshape}). As we mentioned before, the smaller the value of $n$ is, the flatter the potential gets. Therefore, $\lambda\rightarrow0$ in the limit $n\rightarrow0$. In Fig. \ref{fig:lambdaphimodel}, a representation of $\lambda$ as a function of $\phi$ is shown. We realise that, if at early times the scalar field approaches values close to zero, then we will be in a regime where $\lambda\gg1$ and $\Gamma$ will satisfy the conditions needed in order to have tracking behaviour. In addition, as the field evolves towards the minimum, $\lambda$ will tend to $0$ and $\Gamma$ will grow accordingly, thus explaining the late dominance of the dark energy. This is the natural evolution of the scalar field in the model proposed here. In the next section, we will see in detail these mechanisms and how they emerge in the dynamical system approach.    

The expression \eqref{lambdaphimodel} of $\lambda(\phi)$ can be  inverted. Therefore, the dynamical system which we will next study will be well defined. In fact, $\phi$ as a function of $\lambda$ is given by the analytical expression

\begin{equation}\label{philambdamodel}
   \phi(\lambda)=
   \begin{dcases} 
      \eta\arccos{\left(\frac{\lambda^2-n^2/k^2\eta^2}{\lambda^2+n^2/k^2\eta^2}\right)} & \lambda<0, \\
      2\pi\eta-\eta\arccos{\left(\frac{\lambda^2-n^2/k^2\eta^2}{\lambda^2+n^2/k^2\eta^2}\right)} & \lambda\geq 0. \\
   \end{dcases}
\end{equation}

We also note that most of the key analytical results presented in this section, including the structure of the potential and the associated cosmological implications, have been independently derived in recent works \cite{Hossain:2023lxs, Hossain:2025grx}. Their findings are in strong agreement with ours and support the robustness of the potential construction and its role in late-time cosmic acceleration.

\section{A dynamical system approach for our model}\label{dsam}

In this section, we apply a dynamical system approach to our new model and perform numerical calculations in order to see how the tracking behaviour can show up in this model.

In Section \ref{sfdem}, we introduced the scalar field formalism in the presence of a single barotropic fluid with a constant equation of state. In our model, we extend the methodology by introducing two perfect fluids in addition to the scalar field: one corresponding to radiation whose equation of state reads $w_{r}=1/3$, and another one corresponding to dark and baryonic matter with an equation of state $w_{m}=0$. The dynamical system now can be written as

\begin{equation}\label{xfinaleq}
   x'=\frac{1}{2}\left[-3x+3x^3-3xy^2+xz^2\right]+\sqrt{\frac{3}{2}}\lambda y^2,
\end{equation}

\begin{equation}\label{yfinaleq}
   y'=\frac{1}{2}\left[3y-3y^3+3yx^2+yz^2\right]-\sqrt{\frac{3}{2}}\lambda xy,
\end{equation}

\begin{equation}\label{zfinaleq}
   z'=\frac{1}{2}\left[-z+z^3+3zx^2-3zy^2\right],
\end{equation}

\begin{equation}\label{lambdafinaleq}
   \lambda'=-\sqrt{6}f(\lambda)x,
\end{equation}
where $x$ and $y$ are given by \eqref{xexpression} and \eqref{yexpression}, respectively. We have also introduced the new variable $z$, which takes into account the radiation component and is defined as

\begin{equation}\label{zexpression}
   z=\Omega_{r}^{1/2}=\frac{k\sqrt{\rho_r}}{\sqrt{3}H},
\end{equation}
where $\rho_r$ is the energy density of the radiation. The matter component does not appear in the dynamical system because the Friedmann equation \eqref{friedmanneq} is a constraint: $x^2+y^2+z^2+\Omega_{m}=1$, where $\Omega_{m}=k^2\rho_{m}/(3H^2)$ and $\rho_{m}$ is the matter energy density. Again, since $\rho_{m}>0$, the constraint imposes limits in the values of the variables: $0\leq x^2+y^2+z^2\leq1$. In our model, the function $f(\lambda)$ which appears in \eqref{lambdafinaleq} is given by

\begin{equation}\label{flambda}
   f(\lambda)=\lambda^2[\Gamma(\lambda)-1]=\frac{n}{2k^2\eta^2}+\frac{\lambda^2}{2n}.
\end{equation}

\begin{table*}[t]
    \centering
    \begin{tabular}{||c|c|c|c|c|c|c|c|c|c||}
    \hline
       \:Point\: & \:$x$\: & \:\:$y$\:\: & \:$z$\: & $\lambda$ & $w_{\phi}$ & \:$w_{\textrm{eff}}$\: & \:$\Omega_{m}$\: & \:\:$\Omega_{\phi}$\:\: & Stability\\
       \hline
       \hline
       A$_1$  & $0$ & $0$ & $1$ & \:$\lambda_{\star}$\: & \:Undefined\: & $1/3$ & $0$ & $0$ & Saddle\\
       A$_2$  & $0$ & $0$ & $0$ & \:$\lambda_{\star}$\: & \:Undefined\: & $0$ & $1$ & $0$ & Saddle\\
       E  & $0$ & $1$ & $0$ & $0$ & $-1$ & $-1$ & $0$ & $1$ & \:Stable if $n>0$, unstable if $n<0$\:\\
       F$^*$  & \:$x_{\star}$\: & $0$ & \:$z_{\star}$\: & \:Undefined ($\xi=1$)\: & $1$ & \:$x_{\star}^2+z_{\star}^2/3$\: & \:$1-x_{\star}^2-z_{\star}^2$\: & \:$x_{\star}^2$\: & Saddle if $n>0$\\
       \hline
    \end{tabular}
    \caption{\justifying{Fixed points of the system given by \eqref{xfinaleq}, \eqref{yfinaleq}, \eqref{zfinaleq} and \eqref{lambdafinaleq}. F$^*$ is deduced by compactifying the $\lambda$ variable as shown in \eqref{compactlambda}. The quantities $x_{\star}$, $y_{\star}$ and $\lambda_{\star}$ are random values of the variables $x$, $y$ and $\lambda$ inside their domains and which satisfy the constraints detailed in the text (please see the sentence after \eqref{zexpression}).}} 
    \label{tab:fixedpointsmodel}
\end{table*}

Since $\Gamma(\lambda)=\Gamma(-\lambda)$ in our model, the dynamical system remains invariant under the transformations $\lambda\rightarrow-\lambda$, $x\rightarrow-x$. Therefore, the dynamics for negative values of $\lambda$ is analogous and we can restrict our analysis to the positive half domain. We find three fixed points in the finite regime of the phase space. They are classified in Tab. \ref{tab:fixedpointsmodel}. The most important one is the fixed point E which, as we discussed in Section \ref{sfdem}, corresponds to a de Sitter-like universe completely dominated by the scalar field potential. We see that it is an attractor in our model and will be crucial in order to explain the late-time acceleration of our Universe. We found as well two fixed points, A$_{1}$ and A$_2$, which are saddle points and correspond to the radiation and matter eras, respectively.

For completeness, in order to study the dynamical system in the limits $\lambda\rightarrow\pm\infty$, we must compactify the variable $\lambda$. Defining a compactified variable as \cite{Ng:2001hs}

\begin{equation}\label{compactlambda}
   \xi=\frac{\lambda}{1+\lambda},
\end{equation}
the limit $\lambda\rightarrow+\infty$ now corresponds to $\xi=1$. The value $\lambda=0$ is still given by $\xi=0$. We find an extra type of fixed point located at $y=0$, $\xi=1$ for arbitrary $x$ and $z$: for $n>0$, it is a saddle (cf. Tab. \ref{tab:fixedpointsmodel}).

The total equation of state \eqref{effeos} must include the $z$ variable under this generalisation and reads now as

\begin{equation}\label{totaleos}
   w_{\textrm{eff}}=\frac{p_r+p_m+p_{\phi}}{\rho_r+\rho_m+\rho_{\phi}}=x^2-y^2+\frac{1}{3}z^2.
\end{equation}

\begin{figure*}[t]
\begin{subfigure}{0.46\textwidth}
\includegraphics[width=\textwidth]{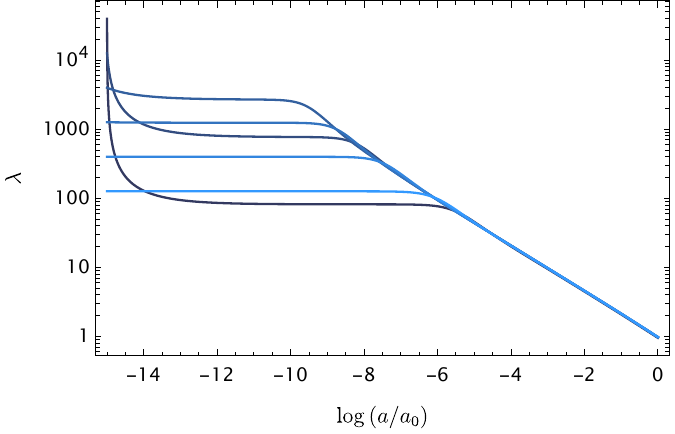}
\caption{Representation of the evolution of $\lambda$ for $n=1$.}
\label{fig:lambda1}    
\end{subfigure}
\hspace{1cm}
\begin{subfigure}{0.46\textwidth}
\includegraphics[width=\textwidth]{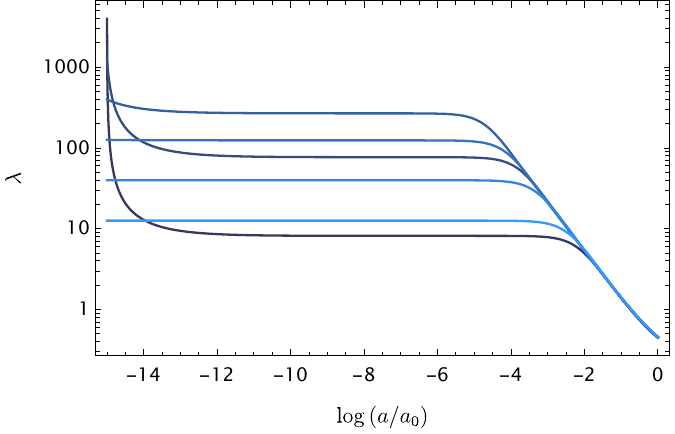}
\caption{Representation of the evolution of $\lambda$ for $n=0.1$.}
\label{fig:lambda01}    
\end{subfigure}
\caption{\justifying{The left hand side panel shows the evolution of the parameter $\lambda$ for $n=1$. Each colour represents a different set of initial conditions in the calculation of the evolution: from $\{\phi_i=5\times10^{-5}M_P, x_i=10^{-2}\}$ (darkest blue) to $\{\phi_i=5\times10^{-5/2}M_P, x_i=10^{-7}\}$ (lightest blue). Note that the initial value $\lambda_i$ can be calculated using \eqref{lambdaphimodel} for each case, and the values of $\dot{\phi_i}$ can be obtained from \eqref{xexpression}. The right hand side panel shows the evolution of $\lambda$ for $n=0.1$. Each colour represents the same set of initial conditions as in the left hand side panel. In both panels, a convergence to a common tracking path is observed, independently of the initial conditions imposed in the calculation.}}
\end{figure*}

\begin{figure*}[t]
\centering
\begin{subfigure}{0.46\textwidth}
\includegraphics[width=\textwidth]{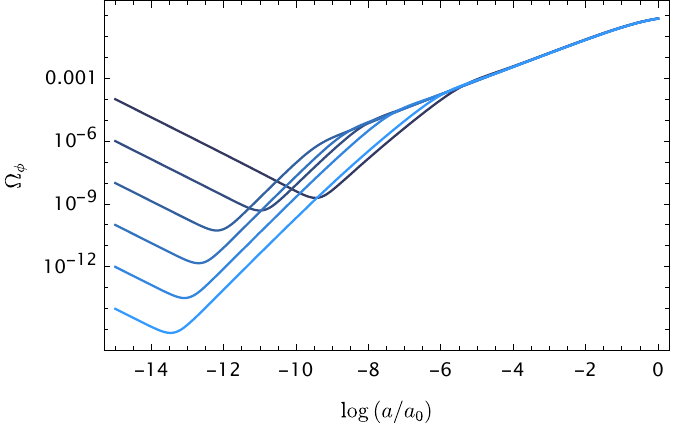}
\caption{Representation of the evolution of $\Omega_{\phi}$ for $n=1$.}
\label{fig:omegaphi1}    
\end{subfigure}
\hspace{1cm}
\begin{subfigure}{0.46\textwidth}
\includegraphics[width=\textwidth]{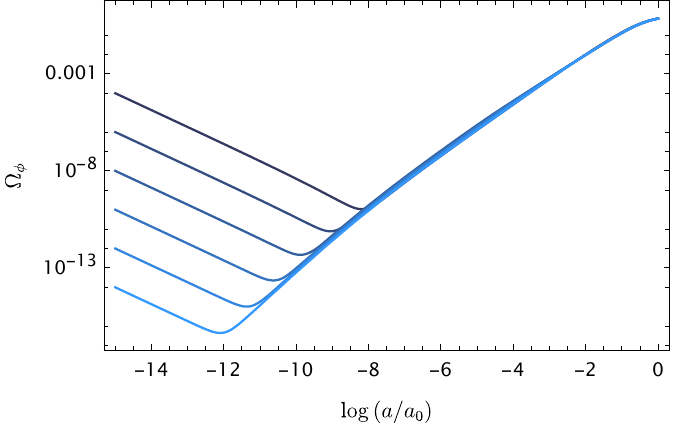}
\caption{Representation of the evolution of $\Omega_{\phi}$ for $n=0.1$.}
\label{fig:omegaphi01}    
\end{subfigure}
\caption{\justifying{The left hand side panel shows the evolution of the scalar field density parameter $\Omega_{\phi}$ for $n=1$. Each colour represents the same set of initial conditions as in Figs. \ref{fig:lambda1} and \ref{fig:lambda01}. A broad range $10^{-14}-10^{-4}$ of initial values $\Omega_{\phi i}$ is included under the sets of initial conditions considered. The right hand side shows the evolution of $\Omega_{\phi}$ for $n=0.1$. In both panels, a convergence to a common tracking path is observed, independently of the initial conditions imposed in the calculation.}}
\end{figure*}

The evolution of the Universe described by our model can be simulated solving numerically the system of equations given by \eqref{xfinaleq}, \eqref{yfinaleq}, \eqref{zfinaleq} and \eqref{lambdafinaleq}. In order to do this, we must first specify the initial conditions of the variables $x$, $y$, $z$ and $\lambda$. We begin our numerical integration at $\log{(a_i/a_0)}=-15$. In order to have an estimation of our initial condition for $z$ around $\log{(a_i/a_0)}=-15$, i.e., during the radiation dominated epoch, we assume that our model is well approximated by a $\Lambda$CDM framework as at that redshift dark energy plays no role. Given this, we calculate the initial value of $z$ as $z_i=\Omega^{1/2}_{r\Lambda i}$, where $\Omega_{r\Lambda i}$ is given by

\begin{equation}\label{omegarcosmologicalconstant}
   \Omega_{r\Lambda i}=\frac{\Omega_{r0}(a_i/a_0)^{-4}}{\Omega_{r0}(a_i/a_0)^{-4}+\Omega_{m0}(a_i/a_0)^{-3}+\Omega_{\Lambda0}},
\end{equation}
and where we use $\Omega_{m0}=0.3$, $\Omega_{r0}=8\times10^{-5}$ and $\Omega_{\Lambda0}=1-\Omega_{m0}-\Omega_{r0}$. As we saw in the previous section, $\lambda(\phi)$ can be inverted in our model. That means that specifying the initial value of $\lambda$ is analogous to specifying the initial value of $\phi$. We perform the numerical calculus for different values of $\phi_i$ in the range $5\times10^{-5}M_P-5\times10^{-5/2}M_P$ in order to see how the convergence of the solutions to the tracking path happens. We have chosen this range of initial values in order to satisfy two key aspects which will be discussed in more detail in the next paragraph when discussing Figs. \ref{fig:lambda1} and \ref{fig:lambda01}: (i) to be initially far from the minimum, $\phi/\eta=\pi$, and (ii) to avoid a frozen regime too close to the minimum. The variable $x$ represents the kinetic part of the field and is related to $\dot{\phi}$ through \eqref{xexpression}. In the same way as for $\lambda$, we study the evolution of the system for different values of $x_i$ in the range $10^{-7}-10^{-2}$ (note that $x$ is dimensionless). The range of initial values chosen for $\phi_i$ and $x_i$ corresponds to a broad range of initial values of $\Omega_{\phi i}$ (cf. Figs. \ref{fig:omegaphi1} and \ref{fig:omegaphi01}). Finally, the variable $y$ is related to the variable $\phi$ through the potential, as can be seen in \eqref{yexpression}; it depends also on $H$. With this, the initial value $y_i$ will be determined by $V(\phi_i)$ and $H_i$. Considering that the Universe is dominated by radiation at early times, $H_i$ is well approximated by $H_i\approx H_0\sqrt{\Omega_{r0}(a_i/a_0)^{-4}}$, where $H_0$ is the Hubble parameter today.

In Figs. \ref{fig:lambda1} and \ref{fig:lambda01}, we have represented the evolution of $\lambda$ for models with $n=1$ and $n=0.1$, respectively. We have done it for several initial conditions (the ones discussed above). We see how the range considered for $\phi_i$ corresponds to a broad range of $\lambda_i$. In fact, the range was taken in order to satisfy two key aspects, as mentioned above. First, we are interested in studying the dynamical evolution of the scalar field throughout different epochs of the Universe. For this reason, the initial value of $\phi_i$ must be small enough to be far from the minimum of the potential at $\phi/\eta=\pi$. This imposes initial conditions satisfying $\phi_i/\eta\ll1$. We remember that the function $\lambda(\phi)$ diverges in the limits $\phi\rightarrow0$, as can be seen in Fig. \ref{fig:lambdaphimodel}. Therefore, the initial values considered for the field impose initial values $\lambda_i$ in the regime $\lambda_i\gg1$. In fact, we see in Figs. \ref{fig:lambda1} and \ref{fig:lambda01} that the field evolves in the regime $\lambda\gg1$ until it approaches nowadays, i.e., $\log{(a/a_0)}=0$. The second aspect to consider is the value of $\lambda$ at which the field is frozen. As we see in Figs. \ref{fig:lambda1} and \ref{fig:lambda01}, the field remains constant for a long period. The frozen field must be far enough from the minimum of the potential in order to join the tracking path before the present time and the model to be distinguishable from $\Lambda$CDM. This imposes upper and lower limits in the range of the initial conditions. For smaller $\phi_i$ (bigger $\lambda_i$), the kinetic energy gained at early times is bigger and the field rapidly evolves to a bigger value (smaller $\lambda$) where it freezes. Therefore, we must be careful when choosing too small values of $\phi_i$ and we should impose a lower limit. On the other hand, when we consider too big values of $\phi_i$ (too small values of $\lambda_i$), the field remains frozen from the beginning. Since the field must leave this regime before $\log{(a/a_0)}=0$ is reached in order to have a non trivial behaviour, this imposes an upper limit in the range of possible initial conditions $\phi_i$. The range proposed before satisfies all the requirements and the field shows the correct behaviour which we were looking for: it reaches the common tracking path before $\log{(a/a_0)}=0$ is reached and after the frozen regime is left. For further discussion on this topics, see \cite{Steinhardt:1999nw}.

\begin{figure*}[t]
\centering
\begin{subfigure}{0.45\textwidth}
\includegraphics[width=\textwidth]{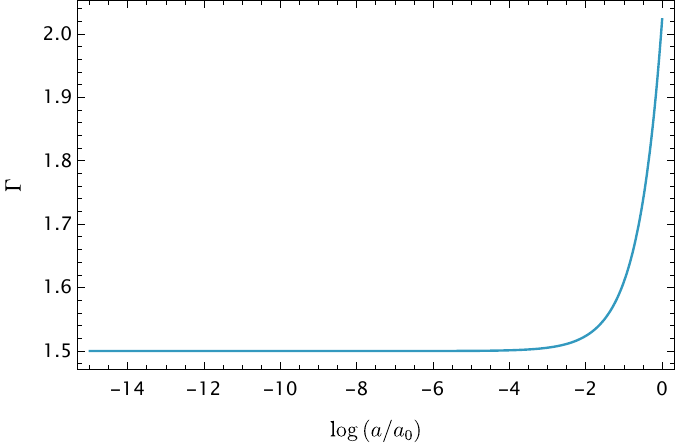}
\caption{Representation of the evolution of $\Gamma$ for $n=1$.}
\label{fig:gamma1}    
\end{subfigure}
\hspace{1cm}
\begin{subfigure}{0.45\textwidth}
\includegraphics[width=\textwidth]{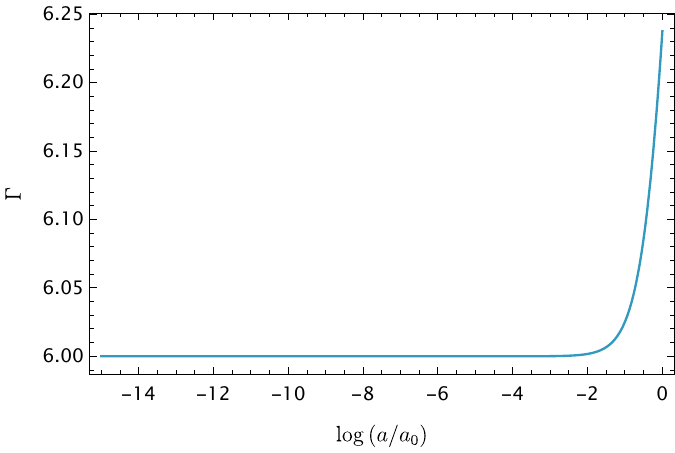}
\caption{Representation of the evolution of $\Gamma$ for $n=0.1$.}
\label{fig:gamma01}    
\end{subfigure}
\caption{\justifying{The left hand side panel shows the evolution of $\Gamma$ for $n=1$. The constant behaviour predicted in \eqref{gammamodelapprox}, responsible of the tracking behaviour, is observed. In this case, $\Gamma\approx3/2$. At late-time, after the convergence to the tracking path, $\Gamma$ grows up and diverges in the limit $\lambda\rightarrow0$, as can be seen in \eqref{gammamodel}. The right hand side panel shows the evolution of $\Gamma$ for $n=0.1$. The same behaviour as in the left hand side panel is observed. In this case, the constant value at early times is $\Gamma\approx6$.}}
\end{figure*}

As we have just discussed, the field evolves in the regime $\lambda\gg1$ for a large amount of time before the present time, i.e., $\log{(a/a_0)}=0$. In fact, only at late-time when the tracking path has already been reached the field evolves towards values where $\lambda\sim1$. In the regime $\lambda\gg1$, \eqref{gammamodel} can be approximated as

\begin{equation}\label{gammamodelapprox}
   \Gamma\approx1+\frac{1}{2n}.
\end{equation}
We find a constant $\Gamma$ which satisfy $\Gamma>1$ in the regime $\lambda\gg1$. Therefore, the conditions of the theorem which ensures tracking behaviour and we presented in Section \ref{sfdem} are satisfied. The evolution of $\lambda$ represented in Figs. \ref{fig:lambda1} and \ref{fig:lambda01} and discussed before is thus the expected one: at some point after the scalar field is frozen, all the paths converge to the common tracking. For $n=1$, we obtain $\Gamma\approx3/2$ and for $n=0.1$, $\Gamma\approx6$. In Figs. \ref{fig:gamma1} and \ref{fig:gamma01}, we have represented the evolution of $\Gamma$ in our model for $n=1$ and $n=0.1$, respectively. We see the expected constant behaviour in the regime $\lambda\gg1$. As the field evolves towards values $\lambda\approx1$, $\Gamma$ loses the constant behaviour. This happens at late-time near $\log{(a/a_0)}=0$: we see that $\Gamma$ grows at these times. In fact, since the late-time attractor corresponding to the fixed point D is located at $\lambda=0$, the function $\Gamma$ tends to diverge. This is precisely the behaviour necessary to naturally explain the late-time dominance of dark energy which we discussed in Section \ref{sfdem}.

\begin{figure*}[t]
\centering
\begin{subfigure}{0.46\textwidth}
\includegraphics[width=\textwidth]{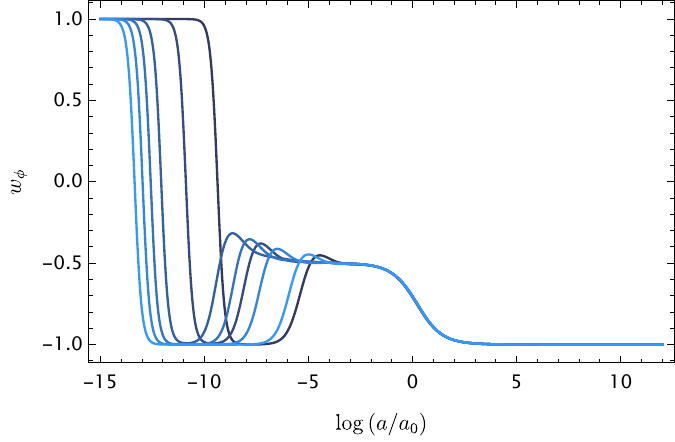}
\caption{Representation of the evolution of $w_{\phi}$ for $n=1$.}
\label{fig:sfeos1}    
\end{subfigure}
\hspace{1cm}
\begin{subfigure}{0.46\textwidth}
\includegraphics[width=\textwidth]{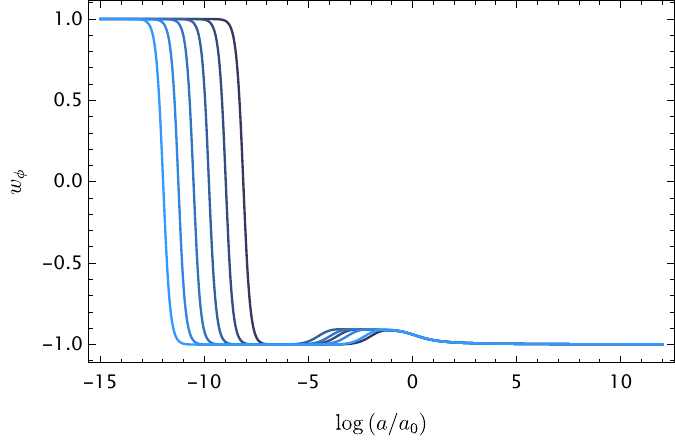}
\caption{Representation of the evolution of $w_{\phi}$ for $n=0.1$.}
\label{fig:sfeos01}    
\end{subfigure}
\caption{\justifying{The left hand side panel shows the evolution of the equation of state of the scalar field $w_{\phi}$ for $n=1$. Each colour represents the same set of initial conditions as in Figs. \ref{fig:lambda1} and \ref{fig:lambda01}.The plot shows an initial kinetic energy dominance ($w_{\phi}\approx1$) which rapidly evolves to a period in which the field remains frozen ($w_{\phi}\approx-1$). After the frozen regime, the field evolves towards the tracking path where $w_{\phi}\approx-1/2$, as expected from \eqref{sfeosgammamodel}. The field finally evolves to the de Sitter attractor correspondent to $\lambda=0$, where $w_{\phi}=-1$. The right hand side panel shows the evolution of the equation of state of the scalar field $w_{\phi}$ for $n=0.1$. The evolution is similar to the one observed in the left hand side panel. In this case, the value of the equation of state in the tracking regime is given by $w_{\phi}\approx-10/11$.}}
\end{figure*}

In Figs. \ref{fig:sfeos1} and \ref{fig:sfeos01} we have represented the evolution of the equation of state of the scalar field in our model for $n=1$ and $n=0.1$, respectively. We see the same tracking behaviour as we saw in Figs. \ref{fig:lambda1} and \ref{fig:lambda01}: at some point, after the freezing regime where $w_{\phi}\approx-1$, all the paths join the common tracking in spite of the early different behaviours showed for different initial conditions. In fact, we can calculate the value of $w_{\phi}$ in the tracking regime by simply introducing the expression \eqref{gammamodelapprox} into \eqref{sfeosgamma}. Since the tracking behaviour is reached at times dominated by the matter component, we can consider the background equation of state $w=0$. We finally obtain

\begin{equation}\label{sfeosgammamodel}
   w_{\phi}\approx-\frac{1}{1+n}.
\end{equation}
If $w_{\phi}$ is calculated for the specific cases studied here, $n=1$ and $n=0.1$, we find $w_{\phi}\approx-1/2$ and $w_{\phi}\approx-10/11$, respectively. We see how the numerical integration shown in the Figs. \ref{fig:sfeos1} and \ref{fig:sfeos01} reproduce our estimations. Note that, in the limit $n\rightarrow0$, we recover $w_{\phi}\rightarrow-1$ in \eqref{sfeosgammamodel} as expected for a cosmological constant. This is in accordance with the representations of Figs. \ref{fig:potentialshape} and \ref{fig:lambdaphimodel}, where we see that the potential is flatter for smaller values of $n$ and, in fact, we recover the $\Lambda$CDM setup in the limit $n\rightarrow0$.

In Figs. \ref{fig:omegaphi1} and \ref{fig:omegaphi01}, we can see how the range of the initial conditions chosen for $\phi$ and $\dot{\phi}$ cover a large range of values of the scalar field density parameter. Although the Universe is dominated by radiation at early times ($\Omega_{ri}\sim1$), we have considered cases where the subdominant components are comparable ($\Omega_{\phi i}\sim\Omega_{mi}$). They are represented in Figs. \ref{fig:omegaphi1} and \ref{fig:omegaphi01} by the biggest values of the density parameter ($\Omega_{\phi i}=10^{-4}$). From that value, we have been gradually lowering the densities towards values closer to that of the cosmological constant. Again, we see that the convergence of the system to the tracking path does not depend on the initial conditions, relaxing the coincidence problem.

\begin{figure*}[t]
\centering
\begin{subfigure}{0.45\textwidth}
\includegraphics[width=\textwidth]{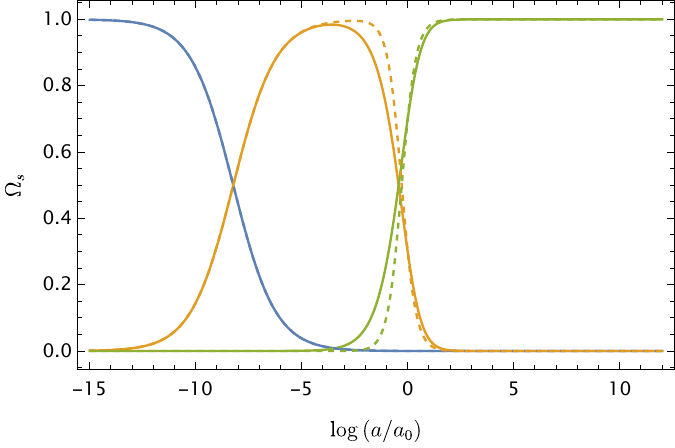}
\caption{Representation of the evolution of density parameters for $n=1$.}
\label{fig:omegas1}    
\end{subfigure}
\hspace{1cm}
\begin{subfigure}{0.45\textwidth}
\includegraphics[width=\textwidth]{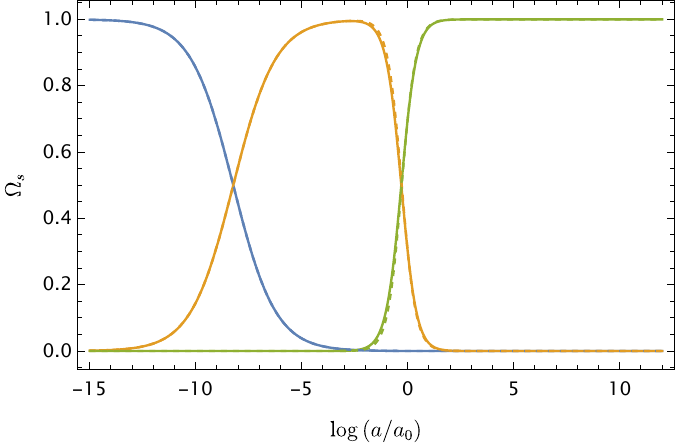}
\caption{Representation of the evolution of density parameters for $n=0.1$.}
\label{fig:omegas01}    
\end{subfigure}
\caption{\justifying{The left hand side panel shows the evolution of the density parameters $\Omega_s$ corresponding to the different components $s$ of the Universe: the radiation density parameter, $s=r$ (blue), the matter density parameter, $s=m$ (yellow), and the dark energy component, $s=\phi$ (green). The continuous lines represent the predictions given by the model proposed in this work for the case of $n=1$, while the discontinuous lines represent the predictions given by a model where the dark energy component is a cosmological constant. The convergence to a common point today has been imposed when performing the numerical calculations. The evolution of the matter and the dark energy components in the model proposed here deviates from the one expected in a cosmological constant model. The right hand side panel shows the analogous evolution of the density parameters for the case of $n=0.1$ (thick lines), comparing it to the cosmological constant one (dashed lines). The discrepancy between the models is notably reduced. The cosmological constant behaviour is reached in the limit $n\rightarrow0$.}}
\end{figure*}

In Figs. \ref{fig:omegas1} and \ref{fig:omegas01}, we have represented the evolution of the density parameters for the models with $n=1$ and $n=0.1$, respectively, and we have compared them to the evolution corresponding to a model where the dark energy is due to a cosmological constant. We see how the evolution is similar for the two cases, but there is an appreciable difference which is crucial in order to explain the matter structure formation. For the first case, $n=1$, in the matter-dominated era $\Omega_{m}$ is a bit smaller at late-time than in the $n=0.1$ case in favour of $\Omega_{\phi}$. This effect is also noticeable in Figs. \ref{fig:omegaphi1} and \ref{fig:omegaphi01}, where we can see how the tracking path of $\Omega_{\phi}$ is located at higher values in the $n=1$ case. This can be understood looking at the expression \eqref{sfeosgammamodel}, which corresponds to the equation of state of the scalar field at the tracking regime. For the case $n=1$, the value of $w_{\phi}$ was found to be further away from the cosmological constant ($w_{v}=-1$) than for the case $n=0.1$. This translates in a greater importance of the scalar field in the matter-dominated era for bigger values of $n$ (further from the cosmological constant), resulting in a smaller decreasing of the values of $\Omega_{m}$. This effect will be important when studying structure formation. In fact, we can not consider arbitrary large values of $n$ in our model since the matter density $\Omega_{m}$ would not be big enough to form structures. This effect is studied in detail in a companion paper where we analyse the perturbations of the model and their implications for structure formation \cite{Boiza:2024fmr}.

\section{Conclusions}\label{conclusions}

In this work, we have first reviewed the theoretical aspects of dark energy models built from a canonical scalar field minimally coupled to gravity, known as quintessence models. We have focused on the dynamical analysis and the mechanisms capable of alleviating the coincidence problem, specially the tracking behaviour, which appears as an attractor in the dynamical system and allows to relax the dependence of the model on the initial conditions. We have then proposed a new $\Gamma$ function (see \eqref{gammadef} for the general definition of $\Gamma$ and \eqref{gammamodel} for the $\Gamma$ proposed in this work) that satisfies the theorem proposed in \cite{Steinhardt:1999nw}, ensuring tracking behaviour with $w_{\phi}<w$ and capable of explaining the late-time acceleration of the Universe through an effective cosmological constant. We have reconstructed the scalar potential associated with the proposed $\Gamma$ function following the mechanism described in \cite{Zhou:2007xp}, and we have found that this potential can be considered a generalisation of the axion-like potential $V(\phi)=\Lambda^4[1-\cos{(\phi/\eta)}]^n$ to negative values of $n$. For this reason, we have called it \textit{generalised axion-like potential}. 

In the second part of the work, we have considered a universe filled with radiation and matter on top of a dark energy component described by a canonical scalar field with the generalised axion-like potential. We have performed a detailed dynamical analysis and we have found three fixed points in the finite regime of the phase space. One of them describes a de Sitter universe and appears as an attractor in the system, capable to explain the late-time acceleration. The other two are saddle points corresponding to matter and radiation. In order to have a non-trivial evolution of the scalar field, we have argued that initially the value of the scalar field must be far from the minimum, which in terms of the dynamical variable $\lambda$ translates into $\lambda\gg 1$ (cf. \eqref{lambdadef} for the general definition of $\lambda$ and \eqref{lambdaphimodel} for the expression of $\lambda$ in our model as a function of $\phi$). In this regime, it has been proven that the conditions of the theorem proposed in \cite{Steinhardt:1999nw} are satisfied, ensuring a tracking regime as long as the frozen regime is not too close to the minimum of the potential. This would ensure a non-trivial evolution of the scalar field that distinguishes our model from $\Lambda$CDM.

Finally, we have reinforced the theoretical discussion with numerical calculations where the evolution of the scalar field can be more easily observed. We have seen how, in the tracking regime, which occurs during the matter-dominated era, all trajectories converge to a common path, and the information about the initial conditions is lost, alleviating the coincidence problem. The numerical calculations have been performed for two cases, $n=1$ and $n=0.1$, after fixing the parameter $\eta$ to $M_P$. We have seen how the evolution of the density parameters is sensitive to the value of the parameter $n$: the higher the value of $n$, the further the equation of state of the scalar field in the tracking regime is from a cosmological constant, which leads to a suppression of $\Omega_m$ that could affect the structure formation. Therefore, we must be careful when choosing the parameter values. 

To complement the theoretical investigation presented here, a companion paper \cite{Chiang:2025qxg} presents a detailed observational analysis of the same generalised axion-like potential. In that work, we confront the model with a suite of up-to-date cosmological datasets, including CMB measurements (Planck 2018 low-$\ell$ TTEE \cite{Planck:2019nip}, NPIPE CamSpec high-$\ell$ TTTEEE \cite{Rosenberg:2022sdy}, and PR4 CMB lensing \cite{Carron:2022eyg,Carron:2022eum}), BAO data from DESI DR1 \cite{DESI:2024mwx,DESI:2024lzq,DESI:2024uvr}, the Pantheon+ Type Ia supernovae compilation \cite{Brout:2022vxf}, low-redshift $H_0$ anchors from Riess et al.~\cite{Riess:2020fzl}, and structure growth data from DES Y1 \cite{DES:2017myr}. The analysis confirms that the model provides a fit to the data comparable to that of $\Lambda$CDM, with no statistically significant improvement nor degradation. Specifically, we find that while the tension between high-redshift (Planck, DESI) and low-redshift (Pantheon+, $H_0$ anchors) datasets is marginally reduced—by about $0.1 \pm 0.3\sigma$—this improvement is offset by an increased tension of $0.3 \pm 0.5\sigma$ with low-$z$ structure growth data from DES Y1. Despite this observational similarity to $\Lambda$CDM, the axion-like model exhibits rich dynamical behavior. Notably, the observationally viable parameter space corresponds to two distinct regimes: one in which the scalar field is heavy and has already settled near the minimum of the potential, and another in which it is so light that it remains effectively frozen until the present day. Both regimes result in background evolution close to $\Lambda$CDM but arise from very different initial conditions and field dynamics. These features may offer useful handles for future model-building, particularly in extended frameworks involving couplings to dark matter, early dark energy components, or multiple fields. We refer the reader to that companion study for detailed statistical results and a comprehensive comparison with the standard cosmological model.

We conclude by mentioning the fifth force which could be present if the scalar field couples to matter. A screening mechanism would be necessary in order to avoid undesirable modifications of gravity in the solar system scale \cite{Khoury:2003aq,Khoury:2003rn,Burrage:2017qrf}. We leave this to a future work.

\newpage

\section*{Acknowledgements}
\begingroup
\setlength{\parskip}{0.2em}  
\setlength{\parsep}{0pt}
The authors are grateful to Hsu-Wen Chiang, Nelson J. Nunes, Tom Broadhurst and Jose Beltrán Jiménez for discussions and insights on the current project. C. G. B. acknowledges financial support from the FPI fellowship PRE2021-100340 of the Spanish Ministry of Science, Innovation and Universities. M. B.-L. is supported by the Basque Foundation of Science Ikerbasque. Our work is supported by the Spanish Grants PID2020-114035GB-100 (MINECO/AEI/FEDER, UE) and PID2023-149016NB-I00 (MINECO/AEI/FEDER, UE). This work is also supported by the Basque government Grant No. IT1628-22 (Spain). 
\endgroup

\bibliography{bibliografia}

\end{document}